\documentclass{aa}
\usepackage{graphicx}
\usepackage[varg]{txfonts}
\usepackage[numbers]{natbib}
\begin{document}

%\draft

\title{Energy dependence of Normal Branch Oscillation in Scorpius X-1}
\titlerunning{Energy dependence of NBO}

\author{J. Wang
          \inst{1}
          \and
          H.-K. Chang
          \inst{1,2}
          \and
          C.-Y. Liu
          \inst{2,3}}

\institute{Institute of Astronomy, National Tsing Hua University,
Hsinchu 30013, Taiwan\\
\email{jwang@mx.nthu.edu.tw}
\and Department of Physics, National Tsing Hua University, Hsinchu
30013, Taiwan\\
\and LESIA, Paris Observatory, 92195 Meudon, France}
%\date{Received date ; accepted date}

%\date{\today}

\abstract {We report the energy dependence of normal branch oscillations (NBOs)
in Scorpius X-1, a low-mass X-ray binary Z-source.
Three characteristic quantities (centroid frequency, quality factor,
and fractional root-mean-squared (rms) amplitude) of a quasi-periodic oscillation signal
as functions of photon energy are investigated.
We found that, although it is not yet statistically well established,
there is a signature indicating that the NBO centroid frequency decreases
with increasing photon energy when it is below 6-8 keV, which
turns out to be positively correlated with the photon energy at the higher energy side.
In addition, the rms amplitude increases significantly with the photon energy below 13 keV
and then decreases
in the energy band of 13-20 keV.
There is no clear dependence on photon energy for the quality factor.
Based on these results, we suggest that the NBO originates mainly
in the transition layer.
 }

\maketitle
%%%%%%%%%%%%%%%%%%%%%%%%%%%%%%%%%%%%%%%%%%%%%%%%%%
\section{Introduction}

%Spectra and timing properties of Sco X-1
The bright persistent neutron star low-mass X-ray binary (LMXB)
Scorpius X-1 traces a "Z" track in the X-ray color-color diagram (CCD),
which consists of three branches --- the horizontal branch (HB),
the normal branch (NB), and the flaring branch (FB, Hasinger \& van der Klis 1989).
The typical Fourier power spectra in Scorpius X-1 are composed of
distinct aperiodic variabilities in most of these branches,
including noise components (broad structure) and
quasi-periodic oscillations (QPOs, narrow feature).
There are three distinct types of QPOs observed
by the Rossi X-Ray Timing Explorer (RXTE), i.e.,
the normal branch oscillation (NBO) at $\sim$ 6 Hz,
the horizontal branch oscillation (HBO) at $\sim$ 45 Hz with a harmonic at
about 90 Hz,
and the kiloherz (kHz) QPOs (van der Klis et al. 1996, 1997).

%description of NBO
The NBOs in Scorpius X-1, which occur on the mid- and lower NB,
show some correlations with the variability of the $\sim$ 45 Hz HBO
and the twin kHz QPOs
(van der Klis et al. 1996; Yu 2007),
and therefore imply some kind of coupling between
the three types of QPOs
(van der Klis et al. 1996; Dieters \& van der Klis 2000).
The NBO frequency remains approximately constant at about 6 Hz, in general.
However, the timing properties along the Z track show that
the NBO frequency extends to at most $\sim$ 21 Hz and
its position moves smoothly into the lower part of the FB
(Priedhorsky et al. 1986; Dieters \& van der Klis 2000).
This smooth transition indicates that
the NBO and flaring branch oscillation (FBO) in Scorpius x-1
are physically related to each other (Kuulkers \& van der Klis 1995; Dieters \& van der Klis 2000; Casella et al. 2006).
Properties of both the twin kHz QPO and 45 Hz HBO depend on the NBO flux.
The upper kHz QPO frequency is anticorrelated with the NBO flux,
and the lower kHz QPO becomes stronger when the NBO flux is low
(Yu et al. 2001).
Significant HBOs are detected during the NBO phase of high flux,
while the HBO disappears at the low flux of NBO phase (Yu 2007).
The coupling between the properties of kHz QPO and HBOs
and the phase of the NBO makes the NBO a unique phenomenon.

%aim and organization
The photons from different regions of the accretion disk
carry different energies and present distinct  physical properties.
So the QPOs detected in different energy band may have varying characteristics.
Inspired by the idea that the photon energy dependence of QPOs
may provide some additional information on the nature of QPOs, and limited
by the available data that provide sufficient spectral information
and at the same time include QPO signals,
we performed a systematical investigation of the energy dependence
of NBOs.
We describe our data reduction and analysis in section 2.
Results are presented in section 3.
In section 4, we discuss the physical implications of our result.
Section 5 contains a short summary.

\section{Data reduction and analysis}

To investigate the energy dependence of QPOs
in Scorpius X-1, high time resolution data with sufficient spectral information
are needed. Quasi-periodic oscillations signals should also be present in these data.
We searched all RXTE observations of Scorpius X-1 for suitable data.
Among these, two binned data modes,
i.e., B\_4ms\_16A\_0\_249\_H and B\_4ms\_16B\_0\_249\_H,
which have a 4-ms time resolution and 16 energy bands consisting
of all PCA channels (from original channel 0 to 249)
binned with two different
schemes (A and B),
provide adequate timing and spectral information for our purpose.
We examined all data recorded in these two modes and selected
those with QPO signals present (with quality factor $Q$ higher than 2)
in the energy band of channel 0-63,
which roughly corresponds to 2 to 20 keV.
The selected data are
listed in Table \ref{tab:obsid}.
We further more defined six energy bands, as shown in Table \ref{tab:c-e},
for the energy dependence study.
An HBO signal was found in the highest energy band defined
in Table \ref{tab:c-e} in the data set 30035-01-07-00,
but it cannot be detected in the original 0-63 channel.
In addition, another $\sim$ 45 Hz HBO signal was detected
in the original 0-63 channel in 93082-01-02-05,
but it cannot be found in any individual energy band defined
in Table \ref{tab:c-e}.
We also found another data set,
30035-01-01-00, which shows the $\sim$ 45 Hz HBO signal
in the original 0-63 channels, but has no NBO.
In a more detailed examination of this data set, its HBO signal
was only detected in the original 0-16 channels but not in higher energy
bands. We therefore did not include this data set in this work.
Limited by the available data, we studied only NBOs.
Some of those data sets include several segments.
We only exploited the segments with an NBO signal
(see Table \ref{tab:obsid} for details).

%In order to understand the state of Scorpius X-1 in each observation,
%we produce the CCDs by using the "Standard 2" mode data.
%The background-subtracted light curves are produced
%with a time resolution of 16 s.
%The soft color is defined as the ratio of count rates
%between (4.7-6.9) keV and (2-4.7) keV,
%and the hard color as that between (9.4-18.2) keV
%and (6.9-9.4) keV.
%Figure \ref{fig:CCD} shows the CCDs for each observation.
%There is no "Standard 2" mode data of ObsID 10056-01-04-02
%in the data archive, so we cannot construct the CCD for this observation.

\begin{figure}
\centering
\includegraphics[bb=30 60 580 800,width=6cm,angle=270]{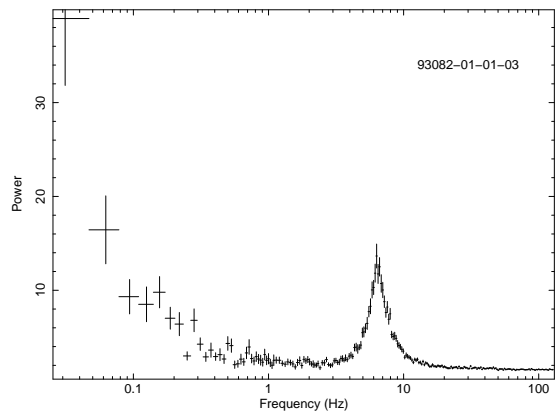}
\caption{One example of PDS which shows a $\sim$ 6 Hz NBO.
The PDS is produced in the PCA original channels 0-63.
The mean count rate is 14348 cts/s/PCU. The exposure time is 938 s.} \label{fig:NBO}
\end{figure}

The data were reduced with the HEASOFT package version 6.8.
Only data with the elevation angle
$> 10^{\circ}$ and the time since South Atlantic Anomaly (SAA)
$<$ 0 or $>$ 30 min were selected for further analysis.
We extracted light curves with a time resolution of 4 ms
($2^{-8}$~s) in the six energy bands defined in Table \ref{tab:c-e}.
%Theses energy bands are so defined that
%there are enough photons in each energy band
%(the mean count rate is still 630 counts/s/PCU in the
%hardest band 36-63 with the lowest statistics)
Using the Leahy normalization (Leahy et al. 1983),
we computed the power density spectra (PDS) without subtracting white noise.
The average power expected from a Poisson distribution is 2.
The original 4 ms ($2^{-8}$~s) time resolution was used,
and each
continuous set was divided into segments of 8192 bins.
A fast Fourier transform (FFT) algorithm for each segment was
computed, yielding a frequency PDS in the range
of $3 \times 10^{-2} - 128$ Hz.
We calculated the average PDS for each continuous data interval.
A logarithmic rebinning of -1.02 was used for the rebin option.
Figure \ref{fig:NBO} shows an example of the $\sim$ 6 Hz NBO signal.

We employed three components to fit the resulting PDS:
a Lorenzian for the very low frequency noise (VLFN),
a power law for the white noise, and another Lorentzian for the NBO.
All fits were performed using the $\chi^2$ minimization technique.
The errors of the analytic model parameters were obtained at
the $\chi^2$ variation of 2.70, which corresponds to
a 90\% confidence for a single parameter.

A QPO signal is characterized by three quantities:
the centroid frequency (representing the position of a QPO),
the quality factor (characterizing the coherence of a QPO signal),
and the fractional root-mean-squared (rms) amplitude
(a measure of the signal strength,
which is proportional to the square root of the peak power contribution
to the PDS). From the  fitting we obtained three parameters of NBOs, i.e.,
the oscillation amplitude, the centroid frequency,
and the full width at half maximum (FWHM).
The quality factor was then computed according to its definition
(centroid frequency/FWHM).
The rms amplitude was calculated by the integral of
the normalized Lorenzian PDS in the appropriate frequency
range and taking the mean count rate of the source into account.
Details of the rms determination can be found
in Lewin et al. (1988) and van der Klis (1989).
The errors of the quality factor and rms amplitude
were estimated with a standard error propagation (Bevington \& Robinson 2003).
%%%%%%%%%Table%%%%%%%%%%%%%%%%%%
\begin{table*}
\begin{center}
\caption{Data log. Here listed are the RXTE observation identification number
(ObsID), their start and stop time, the exposure time in seconds
of the light curve after screening (on elevation and time\_since\_SAA) and
selection of NBO signal presence, the number of working PCU,
the original 0-63 channels count rate in units of cts/s/PCU
and the segments with NBO signals.}
\begin{tabular}{lllllll}
\hline\hline
ObsID \ & \ Start \ & \ End \ & \ Exposure \ & \ PCU \ & \ Rate \ & \ Seg \\
  \ & \   \ & \   \ & \ (s) \ & \   \ & \ (cts/s/PCU) \ & \   \\
\hline\hline
10056-01-04-00 \ & \ 96-05-27  20:42:05 \ & \ 27/05/96  23:02:13 \ & \ 2633 \ & \ 5 \ & \ 18776 \ & \ All \\
\hline
10056-01-04-02 \ & \ 96-05-28  04:26:25 \ & \ 28/05/96  10:29:13 \ & \ 10500 \ & \ 5 \ & \ 17685 \ & \ All \\
\hline
10056-01-05-00 \ & \ 96-05-28  11:11:04 \ & \ 28/05/96  14:23:13 \ & \ 5277 \ & \ 4 \ & \ 16469 \ & \ All \\
 \ & \  \ & \  \ & \ 375 \ & \ 5 \ & \  \ & \  \\
\hline
30035-01-02-000 \ & \ 98-05-31  00:26:47 \ & \ 31/05/98  08:26:47 \ & \ 77 \ & \ 3 \ & \ 18182 \ & \ 4,9 \\
  \ & \  \ & \  \ & \ 2936 \ & \ 5 \ & \  \ & \  \\
\hline
30035-01-05-00 \ & \ 98-06-01  00:26:55 \ & \ 01/06/98  01:18:14 \ & \ 656 \ & \ 3 \ & \ 18413 \ & \ All \\
\hline
30035-01-07-00 \ & \ 98-07-02  06:26:32 \ & \ 02/07/98  10:27:14 \ & \ 4968 \ & \ 5 \ & \ 16486 \ & \ 1,2 \\
\hline
93082-01-01-01 \ & \ 09-06-16  23:43:14 \ & \ 17/06/09  00:56:18 \ & \ 48 \ & \ 4 \ & \ 14718 \ & \ All \\
  \ & \  \ & \  \ & \ 926 \ & \ 5 \ & \  \ & \  \\
\hline
93082-01-01-03 \ & \ 09-06-18  01:40:22 \ & \ 18/06/09  02:21:18 \ & \ 938 \ & \ 5 \ & \ 14348 \ & \ All \\
\hline
93082-01-01-04 \ & \ 09-06-17  05:13:12 \ & \ 17/06/09  06:01:18 \ & \ 1556 \ & \ 5 \ & \ 14486 \ & \ All \\
\hline
93082-01-02-05 \ & \ 09-06-23  04:07:09 \ & \ 23/06/09  04:40:18 \ & \ 1311 \ & \ 3 \ & \ 13447 \ & \ All \\
\hline\hline
\end{tabular}
\label{tab:obsid}
\end{center}
\end{table*}

\begin{table}
\caption{RXTE/PCA energy bands used in this work.}
\begin{center}
\begin{tabular}{l}
\hline
The observations in 1996 and 1998  \\
\end{tabular}
\begin{tabular}{lll}
\hline
PCA \ & \ Energy Band \ & \
Centroid Energy \\
Channel \ & \ (keV) \ & \ (keV) \\
\hline
0-13 \ & \ 1.83-5.07 \ & \ 2.54 \\
14-19 \ & \ 5.07-7.24 \ & \ 6.16 \\
20-22 \ & \ 7.24-8.34 \ & \ 7.79 \\
23-28 \ & \ 8.34-10.52 \ & \ 9.43 \\
29-35 \ & \ 10.52-13.09 \ & \ 11.81 \\
36-63 \ & \ 13.09-23.46 \ & \ 18.26 \\
\hline
\end{tabular}
\begin{tabular}{l}
The observations in 2009  \\
\end{tabular}
\begin{tabular}{lll}
\hline
PCA \ & \ Energy Band \ & \
Centroid Energy \\
Channel \ & \ (keV) \ & \ (keV) \\
\hline
0-11 \ & \ 2.18-5.26  \ &  \ 2.63 \\
12-15 \ & \ 5.26-7.02 \ & \ 6.14 \\
16-19 \ & \ 7.02-8.78 \ & \ 7.90 \\
20-22 \ & \ 8.78-10.10 \ & \ 9.44 \\
23-30 \ & \ 10.10-13.60 \ & \ 11.85 \\
31-63 \ & \ 13.60-28.40 \ & \ 21.00 \\
\hline\hline
\end{tabular}
\label{tab:c-e}
\end{center}
\end{table}

%%%%%%%%%Table%%%%%%%%%%%%%%%%%%

%%%%%%%%%Figure%%%%%%%%%%%%%%%%%%
%%%%%%%%%Figure%%%%%%%%%%%%%%%%%%

\section{Results}

\begin{figure}
\centering
\includegraphics[width=8cm]{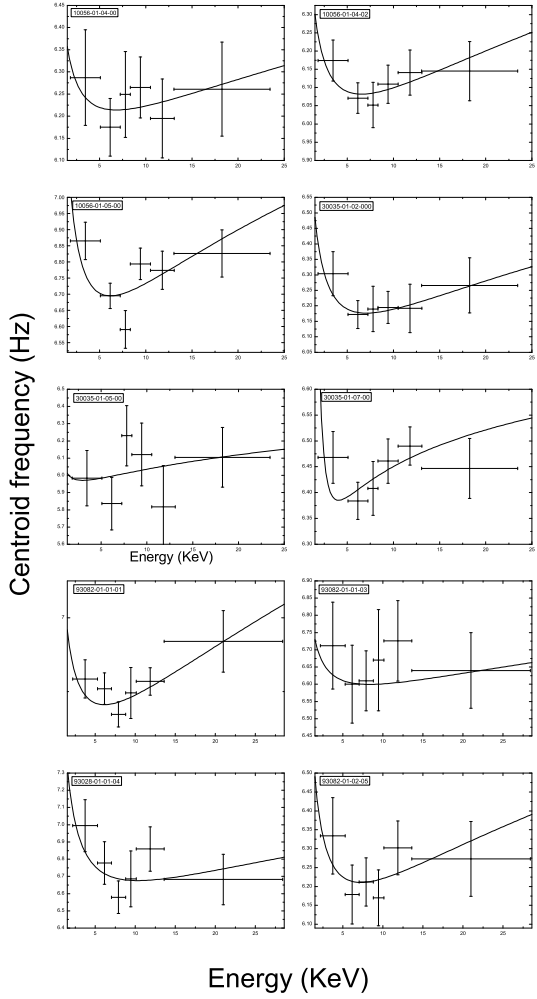}\\
\caption{Centroid frequency versus photon energy
for NBOs in Scorpius X-1. The solid lines are the best fits
with a two-power-law function to the data.}
 \label{fig:fre-e}
\end{figure}

We investigated the photon energy dependence of
three characteristic quantities
(i.e., the centroid frequency, quality factor, and rms amplitude)
of NBOs.
Figure \ref{fig:fre-e} shows the energy dependence of the centroid frequency.
Although error bars are large, there seems to be a common trait
in almost all data sets that the dependence
evolves from a negative correlation to a positive one
with increasing photon energy.
We therefore adopted a two-power-law function,
$\nu(E) = a~E^{\Gamma_1} +~b~E^{\Gamma_2}$,
to describe the relation
between the centroid frequency $\nu$ (in units of Hz)
and the photon energy $E$ (in units of keV).
The fitting results are listed in Table \ref{tab:fre-e-index}.
In all data set we have a negative and a positive power index, which
together cause the centroid frequency to decrease with increasing energy first,
and beyond about 6-8 keV it turns to be positively correlated
with the photon energy.
This dependence is not obvious in one data set (ObsID 30035-01-05-00) among
the ten under study.
While the low $\chi_\nu^2$ value indicates that this description is only
suggestive because of the large error bars,
the appearance of this trait in almost all data sets
asserts its reality.
Fitting this energy dependence with a constant frequency also
gives acceptable results, except for the data set of ObsID
10056-01-05-00. The reduced $\chi^2$ of this constant-frequency
fitting is listed in column 7, denoted with $\chi^2_{\nu,C}$,
 in Table \ref{tab:fre-e-index}.
However, with the poor fitting result of this data set
to  a constant frequency and with the appearance of this `V' shape
trait in almost all data sets, the signature indicating
a `V' shape dependence of the centroid frequency on the photon energy
should not be overlooked. Future observations with better statistics to confirm or
disprove this phenomenon are highly necessary.

\begin{figure}
\centering
\includegraphics[width=8cm]{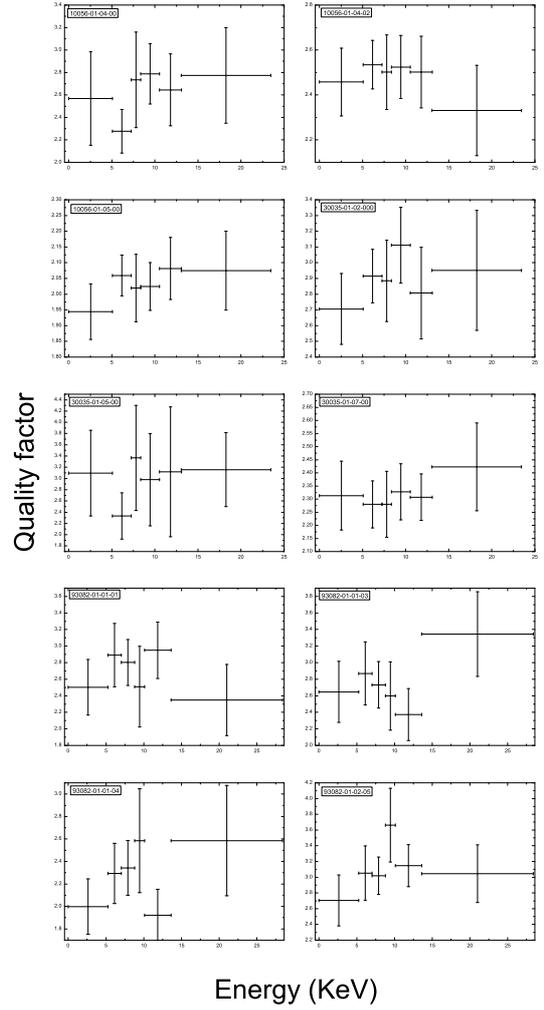}\\
\caption{Quality factor versus photon energy for NBOs in Scorpius X-1.}\label{fig:Q-e}
\end{figure}

Figure \ref{fig:Q-e} shows the energy dependence of the quality factor.
We do not see any clear relation between the quality factor
and the photon energy.
The reduced $\chi^2$ of a constant-quality-factor
fitting is listed in column 8, denoted with $\chi^2_{\nu,Q}$,
in Table \ref{tab:fre-e-index}.
Based on these fitting results,
it is not inconsistent with a constant quality factor.
We note, however, that because of the large error bars, all fittings give
very low reduced $\chi^2$ values.
This means that these data sets are not sufficiently constrained.

\begin{figure}
\centering
\includegraphics[width=8cm]{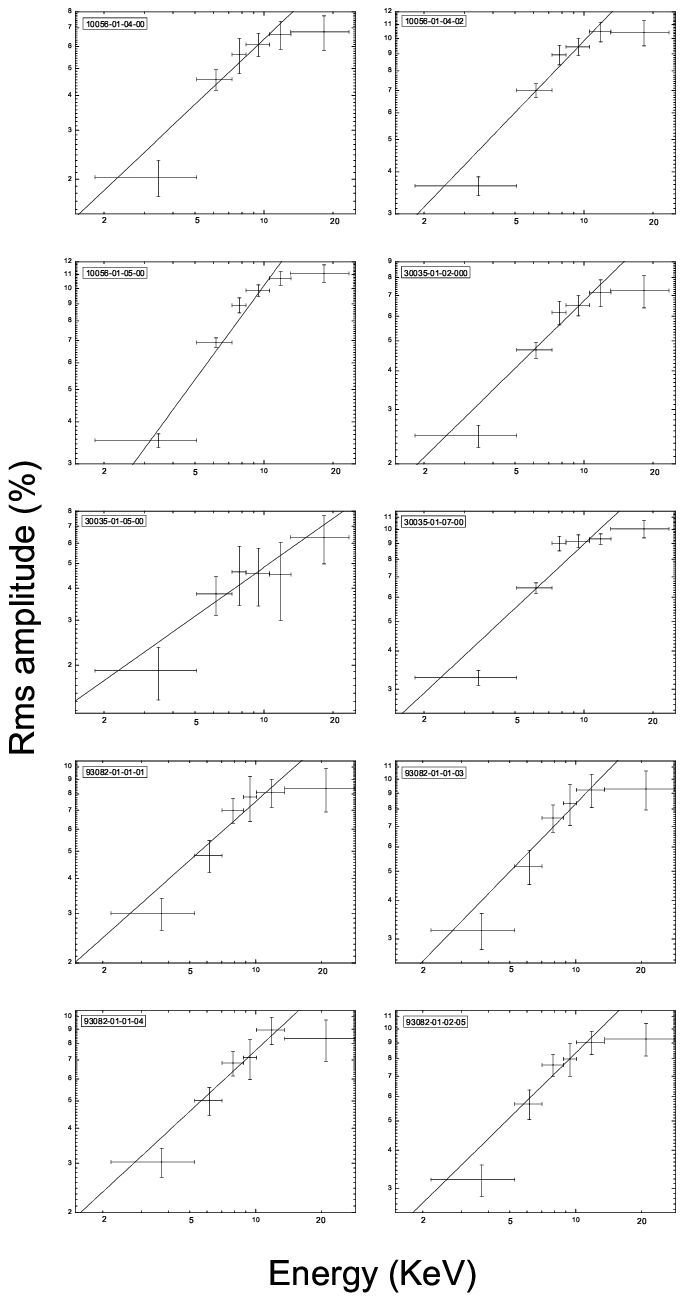}\\
\caption{Rms amplitude versus photon energy for NBOs in Scorpius X-1.
The solid lines are the best fits with a power law to the data points in
the five lower energy bands.}\label{fig:rms-e}
\end{figure}

In Figure \ref{fig:rms-e} we can see that the rms amplitude
increases monotonically with the photon energy
except for the highest energy band.
This means that the strength of the NBO signal increases with the photon energy.
The rms amplitude increases significantly with the photon energy below 13 keV.
However, the increase of rms amplitude with photon energy seems to stop
when the photon energy is higher than $\sim$ 13 keV,
and there might be a plateau.
We do not have enough photons at even higher energies for our study.
It would be interesting to know beyond which energy
the rms amplitude starts to decrease.
To describe the increase of the rms amplitude with photon energy,
we used
a power law relation, rms$(E) = C~E^{\Gamma}$,
to fit the data points in the lower five energy bands.
Table \ref{tab:rms-e-index} shows the fitting results.
Roughly speaking, the rms amplitude increases with photon energy according
to a power law of power index about 0.7 before reaching a possible plateau.
In the data set 30035-01-07-00 the plateau seems to start at a lower energy.

%%%%%%%%%Table%%%%%%%%%%%%%%%%%%
\begin{table*}
\begin{center}
\caption{Fitting results of a
two-power-law description
($\nu(E) = a~E^{\Gamma_1} + b~E^{\Gamma_2}$)
for the energy dependence of the centroid frequency
of NBOs in Scorpius X-1.}
\begin{tabular}{lccccccc}
\hline\hline
ObsID \ & \ $\Gamma_1$ \ & \ a \ & \ $\Gamma_2$ \ & \ b \ & \
$\chi_{\nu}^2$ \ & \ $\chi_{\nu,C}^2$ \ & \ $\chi_{\nu,Q}^2$ \\
\hline\hline
10056-01-04-00 \ & \ -0.17$\pm$0.058 \ & \ 3.40$\pm$0.65 \ & \ 0.11$\pm$0.028 \ & \ 3.05$\pm$0.39 \ & \ 0.19 \ & \ 0.26 \ & \ 0.55 \\
\hline
10056-01-04-02 \ & \ -0.22$\pm$0.026 \ & \ 3.65$\pm$0.29 \ & \ 0.15$\pm$0.030 \ & \ 2.79$\pm$0.43 \ & \ 0.13 \ & \ 0.58 \ & \ 0.15 \\
\hline
10056-01-05-00 \ & \ -0.62$\pm$0.023 \ & \ 2.52$\pm$0.30 \ & \  0.086$\pm$0.025 \ & \ 5.03$\pm$0.38 \ & \ 1.03 \ & \ 2.58 \ & \ 0.26 \\
\hline
30035-01-02-000 \ & \ -0.42$\pm$0.038 \ & \ 2.32$\pm$0.43 \ & \ 0.083$\pm$0.022 \ & \ 4.38$\pm$0.29 \ & \ 0.02 \ & \ 0.51 \ & \ 0.27 \\
\hline
30035-01-05-00 \ & \ -0.43$\pm$0.11 \ & \ 0.81$\pm$1.13 \ & \ 0.040$\pm$0.054 \ & \ 5.23$\pm$0.77 \ & \ 0.65 \ & \ 0.70 \ & \ 0.37 \\
\hline
30035-01-07-00 \ & \ -3.58$\pm$0.27 \ & \ 4.27$\pm$0.30 \ & \ 0.016$\pm$0.014 \ & \ 6.22$\pm$0.20 \ & \ 0.37 \ & \ 0.86 \ & \ 0.11 \\
\hline
93082-01-01-01 \ & \ -0.24$\pm$0.052 \ & \ 5.87$\pm$0.70 \ & \ 0.35$\pm$0.050 \ & \ 1.38$\pm$0.68 \ & \ 0.36 \ & \ 1.17 \ & \ 0.35 \\
\hline
93082-01-01-03 \ & \ -0.16$\pm$0.084 \ & \ 3.16$\pm$1.00 \ & \ 0.082$\pm$0.029 \ & \ 3.65$\pm$0.46 \ & \ 0.15 \ & \ 0.18 \ & \ 0.49 \\
\hline
93082-01-01-04 \ & \ -0.41$\pm$0.057 \ & \ 3.52$\pm$0.85 \ & \ 0.10$\pm$0.042 \ & \ 4.19$\pm$0.68 \ & \ 0.62 \ & \ 1.16 \ & \ 0.65 \\
\hline
93082-01-02-05 \ & \ -0.29$\pm$0.067 \ & \ 3.11$\pm$0.78 \ & \ 0.11$\pm$0.025 \ & \ 3.56$\pm$0.35 \ & \ 0.29 \ & \ 0.57 \ & \ 0.50 \\
\hline\hline
\end{tabular}
\label{tab:fre-e-index}
\end{center}
{\bf Notes.}~~In column 7 ($\chi^2_{\nu,C}$),
listed is the reduced $\chi^2$ of fitting with a constant centroid
frequency.
In column 8 ($\chi^2_{\nu,Q}$),
the reduced $\chi^2$ of fitting the quality factor (Figure \ref{fig:Q-e})
with a constant
is shown.
\end{table*}

\begin{table*}
\begin{center}
\caption{Fitting results of a power law (rms$(E)=C~E^\Gamma$) for
the increase of rms amplitude of NBOs in Scorpius X-1.
Only the five lower energy bands are included in the fitting.}
\begin{tabular}{llll}
\hline\hline
ObsID \ & \ $\Gamma$ \ & \ C \ & \ $\chi_{\nu}^2$ \\
\hline\hline
10056-01-04-00 \ & \ 0.78$\pm$0.18 \ & \ 1.06$\pm$0.40 \ & \ 0.27 \\
\hline
10056-01-04-02 \ & \ 0.71$\pm$0.09 \ & \ 1.92$\pm$0.34 \ & \ 0.44 \\
\hline
10056-01-05-00 \ & \ 0.75$\pm$0.06 \ & \ 1.77$\pm$0.22 \ & \ 0.76 \\
\hline
30035-01-02-000 \ & \ 0.73$\pm$0.12 \ & \ 1.27$\pm$0.30 \ & \ 0.30 \\
\hline
30035-01-05-00 \ & \ 0.63$\pm$0.34 \ & \ 1.12$\pm$0.80 \ & \ 0.14 \\
\hline
30035-01-07-00 \ & \ 0.70$\pm$0.07 \ & \ 1.79$\pm$0.25 \ & \ 3.13 \\
\hline
93082-01-01-01 \ & \ 0.70$\pm$0.19 \ & \ 1.52$\pm$0.61 \ & \ 0.36 \\
\hline
93082-01-01-03 \ & \ 0.74$\pm$0.20 \ & \ 1.52$\pm$0.66 \ & \ 0.28 \\
\hline
93082-01-01-04 \ & \ 0.72$\pm$0.18 \ & \ 1.44$\pm$0.56 \ & \ 0.18 \\
\hline
93082-01-02-05 \ & \ 0.70$\pm$0.16 \ & \ 1.65$\pm$0.55 \ & \ 0.24 \\
\hline\hline
\end{tabular}
\label{tab:rms-e-index}
\end{center}
\end{table*}

%%%%%%%%%Figure%%%%%%%%%%%%%%%%%%
%\begin{figure}
%\centering
%\includegraphics[bb=30 80 600 900,width=2.9cm,angle=270]{10056-01-04-00-fre.ps}%
%\includegraphics[bb=30 80 600 900,width=2.9cm,angle=270]{10056-01-04-02-fre.ps}
%\newline
%\includegraphics[bb=30 80 600 900,width=2.9cm,angle=270]{10056-01-05-00-fre.ps}%
%\includegraphics[bb=30 80 600 900,width=2.9cm,angle=270]{30035-01-02-000-fre.ps}
%\newline
%\includegraphics[bb=30 80 600 900,width=2.9cm,angle=270]{30035-01-05-00-fre.ps}%
%\includegraphics[bb=30 80 600 900,width=2.9cm,angle=270]{30035-01-07-00-fre.ps}
%\newline
%\includegraphics[bb=30 80 600 900,width=2.9cm,angle=270]{93082-01-01-01-fre.ps}%
%\includegraphics[bb=30 80 600 900,width=2.9cm,angle=270]{93082-01-01-03-fre.ps}
%\newline
%\includegraphics[bb=30 80 600 900,width=2.9cm,angle=270]{93082-01-01-04-fre.ps}%
%\includegraphics[bb=30 80 600 900,width=2.9cm,angle=270]{93082-01-02-05-fre.ps}
%\caption{The centroid frequency versus the photon energy
%for NBOs in Scorpius X-1. The solid lines are the best fits
%with a two-power-law function to the data.}
% \label{fig:fre-e}
%\end{figure}

%%%%%%%%%Figure%%%%%%%%%%%%%%%%%%

\section{Discussions}

The mechanism of NBOs was suggested to be related
to the radial oscillation during radiation
feedback (Fortner et al. 1989; Miller \& Park 1995)
or the different modes of disk oscillations
(Alpar et al. 1992; Titarchuk et al. 2001; Bildsten \& Cutler
1995; Bildsten et al. 1996; Hor\'{a}k et al. 2004).
The standard explanation attributes the $\sim$ 6 Hz NBO
to part of the accretion with a near Eddington accretion rate,
in an approximately spherically symmetric radial inflow
at a radius of $\sim$ 100 km
(van der Klis 1995; Hasinger et al. 1990).
According to the radiation hydrodynamic model (Fortner et al. 1989),
NBO originates from a radiation-force/opacity feedback loop
within a spherical flow region at about 300 km from the neutron star surface.
The frequency of the oscillations depends on the luminosity.
Alternatives are the acoustic oscillation in a thick disk
due to the density and optical-thickness perturbations caused
by the rotating medium in a subsonic region of the accretion disk
(the frequency is determined by the rotation rate and vorticity, Alpar et al. 1992), g-mode oscillations arising from the
thermal buoyancy in the ocean on rotating neutron stars
(Bildsten \& Cutler 1995; Bildsten et al. 1996),
the low-frequency modulation in a nonlinear resonance
oscillation of the relativistic disk (Hor\'{a}k et al. 2004),
and the  acoustic oscillations of a spherical shell surface
within $\sim$ 20 km of the neutron star surface
because of the change of accretion geometry in the transition zone
at high accretion rate (Titarchuk et al. 2001).
All of these interpretations focus
on the $\sim$ 6 Hz frequency and its dependence on the accretion rate.
None of them can explain the energy dependence of
the centroid frequency (if it is further confirmed)
and of the rms amplitude.

In the NBOs of Scorpius X-1,
the rms amplitude has a strong energy dependence and the centroid
frequency is also likely to have a `V' shape energy dependence.
The centroid frequency changes from anticorrelation to a positive correlation
at about 6-8 keV.
This nonmonotonic energy dependence of the centroid frequency
implies that the emission zone of the NBOs experiences a
radial variation during the accretion process.
The rms amplitude increases monotonically
and significantly with the photon energy below $\sim$ 13 keV.
When the photon energy is higher than 13 keV,
the rms amplitude seems to stop increasing,
which may indicate that the NBO signals reach the strongest strength at 13-20 keV.
Higher energy photons are often emitted from
the inner region of an accretion disk.
Consequently, the most likely region responsible
for such physics can be referred
to the transition zone between the inner boundary of accretion disk
and the magnetosphere,
in which the transition of the properties for accretion flow occurs
(Titarchuk et al. 1998; Titarchuk \& Osherovich 1999; Titarchuk et al. 1999).
Normal branch oscillation occurs at high accretion rates, i.e., near the Eddington rate.
The high-mass flux deposits increasingly more matter in the transition zone,
leading to the expansion and the radial scale change of this region.
However, the pile of more deposits with greater viscosity can contribute
to the radiation enhancement, as well as to the emission of photons
with higher energies.
As a result, the radial scale change of the transition zone may be
responsible
for the observed nonmonotonic energy dependence of the centroid frequency.
In addition, because of the limited region of the transition zone,
more deposits enhance the viscosity and thus suppress
the further monotonically increasing strength of the NBO signal,
which can explain the energy dependence of rms amplitude at high energy.
We therefore suggest that the NBOs are a type
of oscillations in the transition zone.
This kind of oscillation may originate in the transition layer
due to the viscosity of clumps (e.g. Titarchuk et al. 2001).
On the other hand, this oscillation may also carry the information
of frequency oscillation modes in the accretion disk and
manifest itself as modulation of the accretion rate
during the course when disk matter penetrates the transition layer
(Paczynski 1987; Nowak \& Wagoner 1993; Abramowicz et al. 2007).

\section{Summary}

We investigated the energy dependence of three characteristic quantities of NBOs in Scorpius X-1.
We arrived at the following conclusion:

(1). Although it has yet to be better established statistically,
there is a signature indicating that
the energy dependence of centroid frequency changes from a negative correlation to a positive one.
The turning points are located at 6-8 keV.

(2). It seems that the quality factors have no obvious relation with the photon energy,
which indicates that the coherence of NBOs is independent of the photon energy.

(3). The rms amplitude increases significantly with the photon energy
below 13 keV.
It seems to  reach a plateau beyond that.

(4). The observed energy dependence of three characteristic
quantities of NBOs in Scorpius X-1
{\bf suggests} that the NBOs may be a type of oscillations
in the transition layer.

(5). Horizontal branch oscillation signals are present in three data sets
(30035-01-07-00, 93082-01-02-05, and 30035-01-01-00),
but they display a distinct energy dependence.
The HBO signal is only detected in the original 36-63 channels
for 30035-01-07-00 and in the original 0-16 channels for 30035-01-01-00.
For 93082-01-02-05, the HBO is detected in the original 0-63 channels,
but not in any individual energy band defined in Table \ref{tab:c-e}.
It seems that the energy dependence of HBO signals changes
with the state of the source in Scorpius X-1.
Additional investigation of this phenomenon will be very rewarding
but it requires data of suitable timing and spectral resolution and of sufficiently good
statistics.

\section{Acknowledgments}

This work was supported by the National Science Council of the Republic China under grant
NSC 99-2112-M-007 -017 -MY3.

\end{document}